\documentclass[journal=jctcce,manuscript=article, email-true]{achemso}
\usepackage{amsmath,amssymb,amstext} 
\usepackage{graphicx} 
\usepackage{footmisc} 
\usepackage{bm} 
\usepackage{setspace}
\usepackage{verbatim} 
\usepackage[sort&compress]{natbib}
\SectionNumbersOn
\usepackage{multirow}
\usepackage{array}
\usepackage{longtable}
\usepackage{rotating}
\usepackage{appendix}

\usepackage{titlesec}
\usepackage{chemformula} 
\usepackage[T1]{fontenc} 
\usepackage{siunitx}
\usepackage{subcaption}

\graphicspath{{Pictures/}}
\usepackage{rotating}	
\usepackage{xcolor}
\usepackage{enumitem}

\mciteErrorOnUnknownfalse
\author{Sandra Liz Simon\textit{$^a$}}
\author{Nitin Kaistha{$^{a,}$}}
\email{nkaistha@iitk.ac.in}
\author{Vishal Agarwal\textit{$^{a}$}}
\email{vagarwal@iitk.ac.in}
\affiliation{\textit{$^{a}$}Department of Chemical Engineering, Indian Institute of Technology Kanpur, Kanpur 208016, India.}

\title{Accelerated Relaxation Engines for Optimizing to Minimum Energy Path}

\begin{document}
\begin{abstract}
In the last few decades, several novel algorithms have been designed for  finding critical points on PES and the minimum energy paths connecting them. This has led to considerably improve our understanding of reaction mechanisms and kinetics of the underlying processes. These methods implicitly rely on computation of energy and forces on the PES, which are usually obtained by computationally demanding wave-function or density-function based \emph{ab initio} methods. To mitigate the computational cost, efficient optimization algorithms are needed. Herein, we present two new optimization algorithms: adaptively accelerated relaxation engine (AARE), an enhanced molecular dynamics (MD) scheme, and accelerated conjugate-gradient method (Acc-CG), an improved version of the traditional conjugate gradient (CG) algorithm. We show the efficacy of these algorithms for unconstrained optimization on 2D and 4D test functions. Additionally, we also show the efficacy of these algorithms for optimizing an elastic band of images to the minimum energy path on two analytical potentials (LEPS-I and LEPS-II) and for HCN/CNH isomerization reaction.  In all cases, we find that the new algorithms outperforms the standard and popular fast inertial relaxation engine (FIRE).
\end{abstract}


\clearpage
\section{Introduction}
Heterogeneous catalysis plays a pivotal role in chemical industries and is expected to have a similar impact as we transition to low-carbon technologies.\cite{Sanchez-Bastardo2021} A catalyst with optimal activity, selectivity, and stability is essential to reactor and process design; leading to significant cost reductions and improved process economics. The rational design of the catalyst is achieved through microkinetic modeling,\cite{Motagamwala2021,Chen2021,Jacobsen2002} that breaks down a chemical reaction into elementary steps and constructs model rate equations using critical points on the potential energy surface (PES). The PES describes the energy of a molecule (or a collection of molecules) as a function of its geometry; with local minima representing reactants and products (or intermediates) and the saddle points, which are the `structures of interest' for a chemical reaction. The minimum energy path (MEP), connecting the two neighbouring local minima \emph{via} the saddle point, with the perpendicular component of the energy gradient being zero along this path, outlines the course of the reaction. Therefore, efficient exploration of the PES and finding MEP provides insights into reaction mechanisms and holds great potential for screening and designing active catalysts.

There are several methods for finding the MEP. \cite{Halgren1977,Mller1979,Elber1987,Nichols1990,Fischer1992,Peng1993,Ayala1997,JONSSON1998,Henkelman2000a,Anglada2001,E2002,Schlegel2003,Peters2004,Trygubenko2004,Carr2005,Behn2011}
One of the commonly used methods is the nudged elastic band (NEB) method.\cite{JONSSON1998} NEB is a chain-of-states method where a chain of atomic configurations, referred to as images, is used to discretize the reaction pathway. The images are connected by artificial springs that impose a constraint on the location of the images relative to one another. NEB method employs `nudging', where the perpendicular component of the true force (the negative gradient of potential energy) is combined with the parallel component of the spring force to form the NEB force. An optimization algorithm iteratively guides the images towards MEP, as the gradient of energy is zero along the perpendicular direction of MEP. Therefore, the determination of MEP can be viewed as a constrained optimization problem that requires the NEB forces to be zero. Typically, a discretized NEB pathway consists of five to twelve images, and optimizing the guessed pathway to MEP requires performing several energy and force calculations. When using computationally intensive techniques like \textit{ab initio} methods to obtain energy and forces, the computational effort becomes quite substantial. Therefore, developing methods to reduce this computational burden is a key to efficient exploration of PES. 

One way to alleviate the computational expense associated with the NEB methodology, is to adopt an optimization algorithm, which is robust and efficient. Among the available optimizers, \cite{Sheppard2008,Press1989,Nocedal1980,Bitzek2006,Liu1989} gradient or force-based optimizers such as conjugate gradient (CG), FIRE, L-BFGS are usually preferred as they use the local information provided by the forces for facilitating the relaxation of the elastic band towards MEP. While traditional quasi-Newton methods can also be used, those requiring Hessian evaluation and storage significantly increase computational demands.

In this work, we focus on a force-based optimization technique: fast inertial relaxation engine (FIRE),\cite{Bitzek2006} which employs Newton's equations of motion (see Section \ref{sec:FIRE} for FIRE details). We prioritize FIRE due to its sole dependence on gradient information, thereby eliminating the need to estimate and store the Hessian matrix; a computationally expensive task in quasi-Newton methods. Additionally, FIRE offers the advantage of not requiring computationally expensive line search procedures typically found in conjugate-gradient methods. This aspect is particularly advantageous in scenarios with large system sizes, ensuring competitive computational costs in terms of the number of force evaluations compared to quasi-Newton methods like L-BFGS\cite{Liu1989} and traditional conjugate-gradient search techniques.\cite{Bitzek2006,Ribaldone2022,Guenole2020} While FIRE is efficient, it does exhibit certain limitations. We identify and address these inherent limitations, thereby developing optimization schemes with better convergence characteristics. We assess the performance of these new schemes against FIRE in optimizing a guess to a nearby minima for both 2D and 4D analytical functions. Subsequently, we use them along with NEB methodology to optimize the elastic band to MEP. 

\section{Fast Inertial Relaxation Engine (FIRE) Algorithm} \label{sec:FIRE}
We begin by describing the FIRE algorithm,\cite{Bitzek2006} which is an optimization method used for relaxing the system to a minimum, from an initial guess, using  Newton's equations of motion. 

Consider a system consisting of $N$ nuclei. The total energy  [$E(\mathbf{r}, \mathbf{v})$] of this system will be a function of nuclei coordinates, $\mathbf{r} = \lbrace \mathbf{r}_1, \mathbf{r}_2,\dots,\mathbf{r}_N \rbrace$ and nuclei velocities, $\mathbf{v} = \lbrace \mathbf{v}_1, \mathbf{v}_2,\dots, \mathbf{v}_N \rbrace$. The system experiences forces, which can be obtained by computing the negative gradient of the potential energy, i.e. $\mathbf{F} = -\nabla U(\mathbf{r})$. The minima on the potential energy surface (PES) are the stable states.

In a typical molecular dynamics algorithm, the system is advanced on the PES by numerically integrating Newton's equations of motion. After calculating the new position and velocity at each numerical time-step using numerical integrator, the FIRE algorithm alters the direction and magnitude of velocity based on an adaptive empirical parameter, $\alpha$, and the direction of the current forces. The modified velocity, $\tilde{\mathbf{v}}(t)$, is given by          
\begin{equation}\label{eqn:decV}
\tilde{\mathbf{v}}(t) = (1 - \alpha) \mathbf{v}(t) + \alpha |\mathbf{v}(t)| \hat{\mathbf{F}}(t).
\end{equation}
where  $|\mathbf{v}(t)|$ is the magnitude of velocity at time $t$, and the hat represents the unit vector. When $\alpha = 0$, there is no velocity modification as $\tilde{\mathbf{v}}(t)=\mathbf{v}(t)$. When $\alpha = 1$, the velocities are aligned along the force direction. Starting from an initial value of $\alpha$, the algorithm adaptively makes slight changes to $\alpha$ based on the power factor, $P = \mathbf{F}\cdot\mathbf{v}$. When the system is moving downhill, i.e.\ $P$ is positive as system is moving in the direction of decreasing forces, $\alpha$ is adjusted slightly to continue this motion. Conversely, when the system is moving uphill, i.e.\ $P$ is negative, the algorithm applies a sudden brake, reducing the velocity to zero. 

\begin{figure}[!h]
        \centering
        \includegraphics[width=0.4\textwidth]{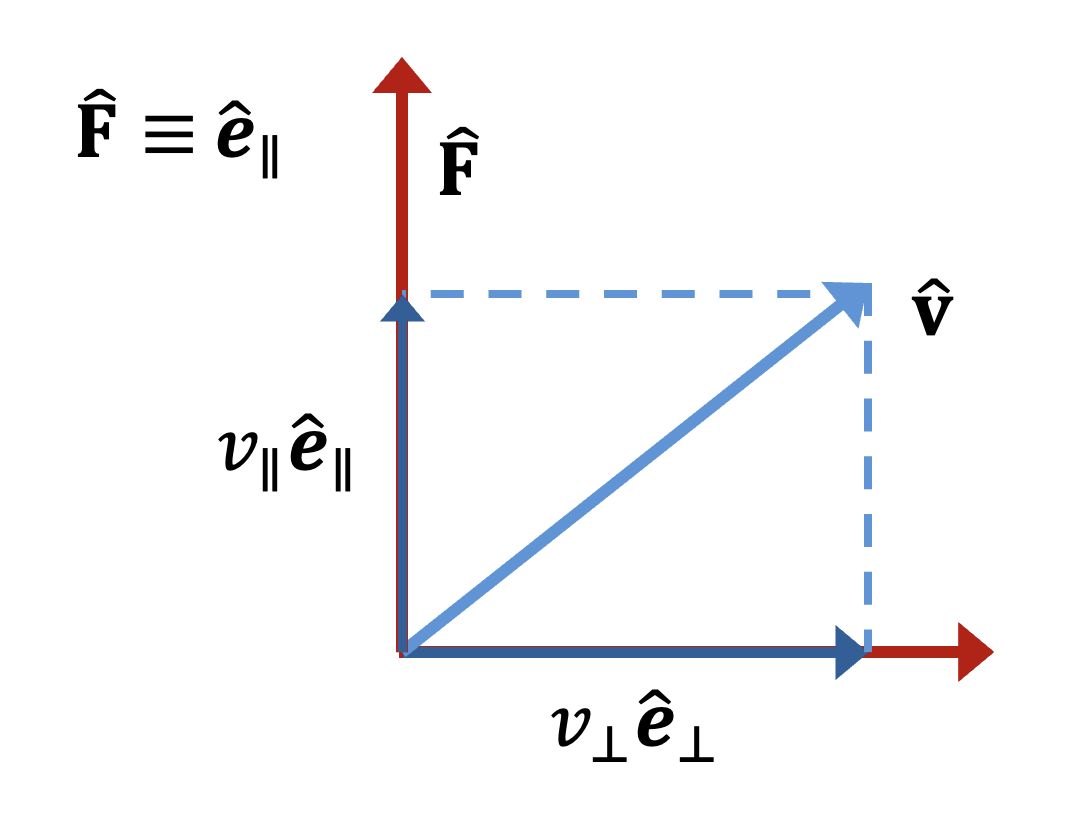}
        \caption{Decomposition of $\hat{\mathbf{v}}$ into components parallel ($v_{\parallel}\mathbf{\hat{e}}_{\parallel}$) and perpendicular ($v_{\perp}\mathbf{\hat{e}}_{\perp}$) to the $\hat{\mathbf{F}}$. Here, $\hat{\mathbf{F}}$ is aligned along the $y$-axis and $|v_{\parallel}|^2 + |v_{\perp}|^2 = 1$. For this analysis, we assume a $two$-dimensional system and the power factor greater than zero. }
        \label{fig:velDecFor}
\end{figure}

For a system moving downhill, the effect of $\alpha$ parameter can be understood by considering a two-dimensional system. To simplify the analysis, we orient the forces on this system along the $y$-axis as shown in Figure \ref{fig:velDecFor}. Further, we decompose the unit velocity vector as $\hat{\mathbf{v}} = v_{\parallel}\hat{\mathbf{e}}_{\parallel} + v_{\perp}\hat{\mathbf{e}}_{\perp}$, where $\mathbf{\hat{e}}_{\parallel}$ and $\mathbf{\hat{e}}_{\perp}$ are the unit vectors parallel and perpendicular to $\hat{\mathbf{F}}$, respectively. The FIRE modified-velocity in Eqn.\ \ref{eqn:decV} for this system can be written as:
\begin{equation}
    \tilde{\mathbf{v}}(t) = |\mathbf{v}(t)|\left[(1-\alpha)v_{\perp}(t)\hat{\mathbf{e}}_{\perp} + \left\lbrace v_{\parallel}(t)\hat{\mathbf{e}}_{\parallel}+\alpha\left[\hat{\mathbf{F}}(t) - v_{\parallel}(t)\hat{\mathbf{e}}_{\parallel}\right]\right\rbrace\right].
\end{equation}
Here $|\mathbf{v}(t)|$ is the magnitude of velocity. Because $\alpha\in$ $[0,1]$ and the quantity $(\hat{\mathbf{F}} - v_{\parallel}\hat{\mathbf{e}}_{\parallel})$ is always positive in downhill motion, the system is accelerated in the force direction and decelerated in the direction(s) perpendicular to the force. FIRE also adjusts the MD integration time-step dynamically according to the power factor. If the power factor is positive $(P>0)$, the time-step is increased and the system takes longer strides. The time-step is increased until reaching the maximum stable time-step ($\Delta t_{max}$) or experiences uphill motion in the current direction ($P<0$). If the power factor is negative ($P<0$), the time-step is reduced by half. In addition, a brief latency period of $N_{min}$ MD steps is included before increasing the time-step to ensure dynamical stability.

For completeness, we provide the FIRE algorithm below:

\begin{enumerate}[align=left]
  \item[\textbf{Step 1:}]
  Begin:
  \begin{itemize}
    \item Set FIRE parameters: initial value of mixing parameter $\alpha_{\text{start}} = 0.1 $; decrement factor for mixing parameter $f_{\alpha} = 0.99 $;  increment factor for time-step $f_{\text{inc}} = 1.1 $; decrement factor for time-step $f_{\text{dec}} = 0.5 $; initial time-step $\Delta t_{\text{start}}= 0.1 \mathrm{fs} $; maximum time-step $ \Delta t_{\text{max}}=10 \Delta t_{\text{start}} $ and latency time $ N_{\text{min}} = 5 $.
    \item Initialize: time $t=0$; position $\mathbf{r} = \mathbf{r}_{\text{start}}$; velocity $\mathbf{v} = 0$; time-step $\Delta t = \Delta t_{\text{start}}$; mixing parameter $\alpha = \alpha_{\text{start}} $ and latency time counter $N_P = 0$.
  \end{itemize}

  \item[\textbf{Step 2:}]
  Compute the forces $\mathbf{F}[\mathbf{r}(t)] = -\nabla U[\mathbf{r}(t)]$. Check for Convergence. If converged, exit the program.

  \item[\textbf{Step 3:}]
  Evaluate the power factor, $P(t) = \mathbf{F}(t) \cdot \mathbf{v}(t)$.

  \item[\textbf{Step 4:}]
  Modify the velocity $\mathbf{v}(t) \rightarrow (1 - \alpha) \mathbf{v}(t) + \alpha |\mathbf{v}(t)| \hat{\mathbf{F}}(t)$.

  \item[\textbf{Step 5:}]
  If $P(t) > 0$, $N_{P} \rightarrow N_{P} + 1$. If $N_{P} > N_{\text{min}}$,
  \begin{itemize}
    \item Increase the time-step, $\Delta t \rightarrow \min(\Delta t \cdot f_{\text{inc}}, \Delta t_{\text{max}})$.
    \item Decrease $\alpha \rightarrow \alpha \cdot f_{\alpha}$.
  \end{itemize}

  \item[\textbf{Step 6:}]
  If $P(t) \leq 0$,
  \begin{itemize}
    \item Decrease the time-step, $\Delta t \rightarrow \Delta t \cdot f_{\text{dec}}$.
    \item Apply brakes, $\mathbf{v}(t) \rightarrow 0$.
    \item Reset $\alpha \rightarrow \alpha_{\text{start}}$.
    \item $N_{P} \rightarrow 0$.
  \end{itemize}

  \item[\textbf{Step 7:}]
  Calculate $\mathbf{r}(t+\Delta{t})$, $\mathbf{v}(t+\Delta{t})$ using an MD integrator. Go to Step 2.
\end{enumerate}

\section{Shortcomings of FIRE}
FIRE is an efficient optimisation scheme, however, it exhibits certain limitations. The algorithm relies on an empirical parameter $\alpha$ for calculating the direction and magnitude of velocities in FIRE. The value of this parameter is set during the start of the program and is adaptively altered based on the uphill or downhill motion of the system. The original FIRE algorithm suggested a value of $\alpha_\text{start}=0.1$. The small value was suggested, may be, because the integrated velocities from molecular dynamics carry information about the nature of the surface. A single value of $\alpha$, however, may not be an optimum for different PES. To verify this, we performed tests on several 2D functions, taken from literature, using the FIRE algorithm. The list of these 2D test functions is provided in Supplementary Information (Table S1). In most of the cases, we find that the optimum value of $\alpha_\text{start}$ (denoted as `$\alpha_\text{start}^\text{opt}$') that minimizes the number of force evaluations to find the minima differs from 0.1 (see Table S2 in Supplementary Information). In several of these cases, the use of an optimum $\alpha_\text{start}$ significantly reduced the number of iterations. However, there is no one value of $\alpha_\text{start}^\text{opt}$ that minimizes the number of iterations and, as expected, is dependent on the nature of the PES. This implies that for each PES, one needs to optimize the value of $\alpha_\text{start}$. The effort done to optimize $\alpha_\text{start}$ will, however, reduce the efficiency of the FIRE algorithm. Therefore, we propose an algorithm (\textit{vide infra}), which is not dependent on the mixing parameter $\alpha$. 

\begin{figure}[!h]
    \centering
    \includegraphics[width=1\linewidth]{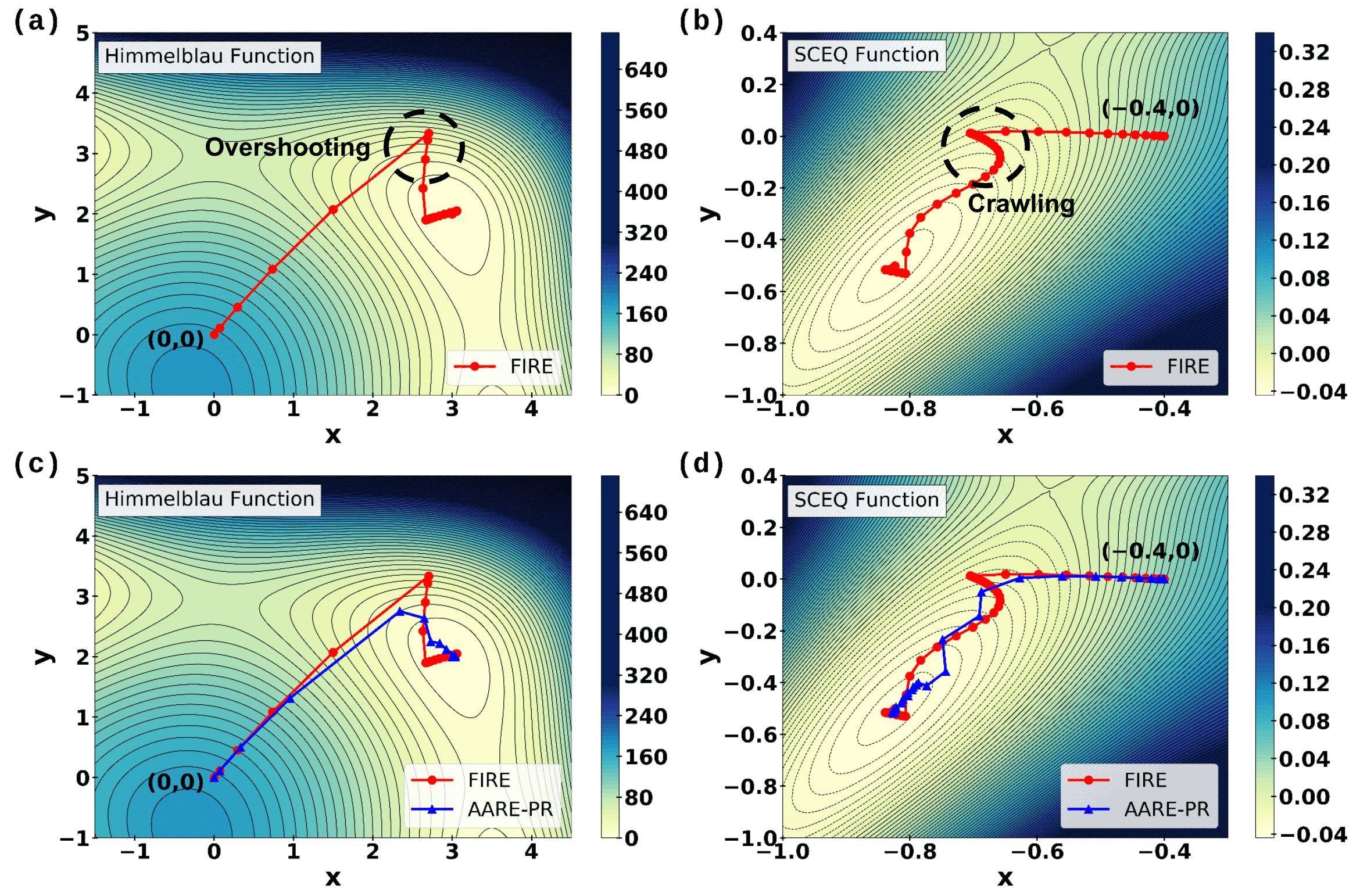}
     \caption{Trajectories of FIRE and AARE on 2D test functions. The mathematical formulas for these functions are provided in Supplementary Information (Table S1). \textbf{(a)} and \textbf{(b)} demonstrates the problems with the FIRE algorithm. In \textbf{(a)} and \textbf{(c)}, we use  the Himmelblau function: starting point $ = \left(0,0\right)$; minima $= \left(3, 2\right)$; and convergence criteria taken as $\|\mathbf{F}\| < 0.01$. In \textbf{(b)} and \textbf{(d)}, we use the SCEQ function: starting point $= \left(-0.4,0\right)$; minima $= \left(-0.83, -0.52\right)$; and convergence criteria taken as $\|\mathbf{F}\| < 0.001$.}
    \label{fig:limitation}
\end{figure}

The other shortcomings of FIRE can be understood by analysing the trajectories on the 2D test functions. Figure \ref{fig:limitation}a shows that a large numerical time-step, as a result of FIRE acceleration, causes the trajectory to ``overshoot'' (i.e.\ the power factor becomes negative). The trajectory spends several iterations after overshooting to recover back to the path of optimization. Additionally, on sensing the uphill motion, FIRE quenches the velocities to zero, decreases the integration time step, and imposes a latency time before resuming the accelerated dynamics. This results in ``crawling'' of the trajectory on the PES, as shown in Figure \ref{fig:limitation}b. A couple of variants of the FIRE algorithm try to address these shortcomings.\cite{Guenole2020,EcheverriRestrepo2023} For instance, Fire 2.0\cite{Guenole2020} proposes to reduce overshooting by sending trajectory backward by half a time step. ABC-FIRE\cite{EcheverriRestrepo2023} introduces a scaling term to correct the bias towards zero velocities on overshooting. However, there is no systematic approach to addressing all the issues with FIRE.

Therefore, we introduce new algorithms that address the aforementioned issues and have the potential to outperform FIRE in terms of the number of force evaluations required. Before presenting these new algorithms, we will discuss their key features and the motivation behind each modification.

\section{Adaptively Accelerated Relaxation Engine (AARE)}

\subsection{Velocity Modification}
One of the critical aspects of our new algorithm is determining the direction of velocity. Instead of relying on an empirical mixing parameter for the determination of direction, as in FIRE, we propose to do away with this parameter. One way to achieve this is by using a value of \emph{one} for this parameter $\alpha$ in the FIRE routine. The modified velocity is given by Eqn. \ref{eqn:firesd} where we move only along the force or steepest-descent (SD) direction with the magnitude of velocities determined by the molecular dynamics algorithm. 
\begin{equation}\label{eqn:firesd}
\tilde{\mathbf{v}}(t) = |\mathbf{v}(t)| \hat{\mathbf{F}}(t).
\end{equation}

Using this velocity modification, we create a FIRE variant referred to as ``FIRE-SD''. We assess how well FIRE-SD performs in comparison to FIRE for optimisation on \emph{twenty-six} 2-dimensional test functions, which are listed in Supplementary Information (see Table S1). As shown in Table \ref{tab:FIRE_comp}, we find that FIRE-SD performs better than the FIRE for most of the cases in terms of the number of force evaluations needed by each algorithm to reach minima. 

Similarly, one can imagine modifying the velocity to the conjugate-gradient (CG) direction rather than the force or steepest descent (SD) direction, as CG directions generally have faster convergence than SD.\cite{chongoptimization} The traditional CG algorithm constructs the direction of descent, $\mathbf{d}_{k}$, using the forces, which are conjugate with respect to the Hessian matrix ($\mathbf{H}$). With the initial descent along the forces, the direction in the $k$-th step is given as
\begin{equation}
\mathbf{d}_{k}  = \mathbf{F}_{k} + \beta_{k} \mathbf{d}_{k-1},
\end{equation}
\begin{equation}
\beta_{k}  = \frac{\left(\mathbf{F}_{k}\right)^\text{T} \mathbf{H}_{k} \mathbf{d}_{k-1}}{\left(\mathbf{d}_{k-1}\right)^\text{T}  \mathbf{H}_{k} \mathbf{d}_{k-1}}.
\end{equation}
where, $\mathbf{F}$ is the negative gradient of the potential energy function, and the superscript `$\text{T}$' denotes the transpose of a vector. Setting the mixing parameter of FIRE, i.e. $\alpha$, to 1 and using unit vector along $\mathbf{d}_{k}$ allows the velocity to be aligned along the CG-direction, yielding the modified velocity equation that follows
\begin{equation}\label{eqn:firecg}
\tilde{\mathbf{v}}(t) = |\mathbf{v}| \cdot \frac{\mathbf{d}_k}{|\mathbf{d}_k|}.
\end{equation}
We denote this algorithm as ``FIRE-CG''. On using the CG direction determined by Polak-Ribi{\`e}re (PR) method (see below for more discussion on this method), as given by Equation \ref{eqn:firecg}, we find that in all the cases, except one, FIRE-PR performs better or is on par with the FIRE and FIRE-SD routines (see Table \ref{tab:FIRE_comp}). This is encouraging; however, we search below for an adaptive direction change algorithm with much better convergence characteristics. 

\begin{table}[!h]
\centering
\caption{Number of force evaluations needed by FIRE, FIRE-SD, and FIRE-PR for finding minima on 2D test functions. The performance ratios relative to FIRE are also tabulated. The mathematical formulas for these functions are provided in Supplementary Information (Table S1). In FIRE-SD and FIRE-PR, the initial value of the mixing parameter $\alpha_{\text{start}} = 1$, and the velocity is aligned along the steepest-descent (SD) and Polak-Ribi{\`e}re (PR) directions; respectively. The iterative loop was terminated when $\|\mathbf{F}\| < 0.01$.}
\label{tab:FIRE_comp}
 \centering
 \resizebox{\textwidth}{!}{%
 \begin{tabular}{@{\extracolsep{\fill}}lc c c c c c c}
 \hline
 \hline
\textbf{Test Function} & \textbf{Starting Point} & \textbf{FIRE} & \textbf{FIRE-SD} & \textbf{$\frac{\textbf{FIRE}}{\textbf{FIRE-SD}}$} & \textbf{FIRE-PR} & \textbf{$\frac{\textbf{FIRE}}{\textbf{FIRE-PR}}$}\\
\hline
Himmelblau Function & (0,0) & 84 & 46 & 1.82 & 41 & 2.05\\
Goldstein $-$ Price Function & (-1,-1) & 47 & 30 & 1.57 & 17 & 2.76 \\
Extended Beale Function & (0,0) & 159 & 138 & 1.15 & 113 & 1.4\\
Rosenbrock Function & (-1.2,1) & 1565 & 2601 & 0.6 & 1230 & 1.27\\
Hager Function & (1,1) & 26 & 26 & 1 & 26 & 1 \\
Booth Function & (0,-5) & 84 & 53 & 1.58 & 57 & 1.47\\
Raydan 1 Function & (3,2) & 38 & 33 & 1.15 & 33 & 1.15\\
Extended Penalty Function & (1,2) & 32 & 32 & 1 & 31 & 1.03\\
Diagonal 1 Function & (0.5,0.5) & 13 & 13 & 1 & 13 & 1\\
Diagonal 2 Function & (1,0.5) & 41 & 31 & 1.32 & 30 & 1.36\\
Diagonal 3 Function & (1,1) & 36 & 27 & 1.33 & 26 & 1.38\\
Tridiagonal 1 Function & (2,2) & 49 & 21 & 2.33 & 21 & 2.33\\
Extended TET Function & (0.1,0.1) & 37 & 22 & 1.68 & 29 & 1.27\\
Generalized PSC1 Function & (3,0.1) & 40 & 22 & 1.82 & 25 & 1.6\\
Full Hessian FH2 Function & (0.01,0.01) & 77 & 40 & 1.92 & 47 & 1.64 \\
Extended BD1 Function & (0.1,0.1) & 30 & 33 & 0.9 & 44 & 0.68\\
Extended Maratos Function & (1.1,0.1) & 1616 & 2535 & 0.64 & 1319  & 1.22\\
Quadratic QF1 Function & (1,1) & 39 & 38 & 1.02 & 28 & 1.39\\
Perturbed Quadratic Function & (0.5,0.5) & 36 & 36 & 1 & 36 & 1 \\
Fletcher Function & (0,0) & 57 & 57 & 1 & 57 & 1  \\
Tridia Function & (1,1) & 54 & 47 & 1.15 & 35 & 1.54\\
Arwhead Function & (1,1) & 70 & 31 & 2.26 & 28  & 2.5\\
EG2 Function & (1,1) & 42 & 38 & 1.1 & 29 & 1.45\\
Liarwhd Function & (4,4) & 126 & 47 & 2.68 & 51 & 2.47 \\
Power Function & (1,1) & 53 & 33 & 1.6 & 32 & 1.65\\
Engval 1 Function & (2,2) & 67 & 29 & 2.31 & 30 & 2.23\\
\hline
\end{tabular}}
\end{table}

When extending CG to handle general non-linear functions, the computational cost escalates due to the need to re-evaluate the Hessian matrix at each iteration. Several variants have been developed to mitigate this problem, and few have survived the rigorous testing over the years. Among these are Hesteness-Stiefel (HS), Polak-Ribi{\`e}re (PR), and Fletcher-Reeves (FR) methods;\cite{Hestenes1952,polak1969,Powell1986,Fletcher1964,Fletcher2000} each offering different ways for calculating the conjugate direction parameter $\beta$ without Hessian evaluation.\cite{Hestenes1952,polak1969,Powell1986,Fletcher1964,Fletcher2000} The $\beta$ parameter for the three cases are given as:
\begin{eqnarray}
\text{\textbf{Hesteness-Stiefel:}} \quad\quad& \beta_{k}  = \frac{\left(\mathbf{F}_{k}\right)^\text{T}\left(\mathbf{F}_{k} -\mathbf{F}_{k-1}\right)}{\left(\mathbf{d}_{k-1}\right)^\text{T}  \left(\mathbf{F}_{k} -\mathbf{F}_{k-1}\right)},   \\
\text{\textbf{Polak-Ribi{\`e}re:}} \quad\quad& \beta_{k}  = \frac{\left(\mathbf{F}_{k}\right)^\text{T}\left(\mathbf{F}_{k} -\mathbf{F}_{k-1}\right)}{\left(\mathbf{F}_{k-1}\right)^\text{T}  \mathbf{F}_{k-1}},  \\
\text{\textbf{Fletcher-Reeves:}} \quad\quad& \beta_{k}  = \frac{\left(\mathbf{F}_{k}\right)^\text{T}\mathbf{F}_{k}}{\left(\mathbf{F}_{k-1}\right)^\text{T}  \mathbf{F}_{k-1}}. 
\end{eqnarray}
All three variants are equivalent for a quadratic surface when an exact line search is used to determine the step size. However, without performing an exact line search and with the velocity direction changing at each step, the direction vectors can only be described as ``CG-like''. As a result, the performance of the Hestenes-Stiefel (HS), Polak-Ribi{\`e}re (PR), and Fletcher-Reeves (FR) differ significantly. The choice between these variants depends on the behavior of the objective function; in some cases, FR performs better, while in others, PR is superior. These variations differ in stability and convergence speed, necessitating the selection of the appropriate variant based on the objective function. 

Given that the choice of CG variant for calculating the $\beta$ parameter is function-dependent, an adaptive direction selection strategy that combines these variants can help leverage the benefits of each. To develop the selection strategy, we begin by analyzing the directional behavior of these variants for a simple quadratic potential. We focus on the angle $\Theta$ between the instantaneous forces ($\mathbf{F}$) and the incoming velocity direction ($\mathbf{d}_{k-1}$), similar to the power factor $P = \mathbf{F}\cdot\mathbf{v}$ in the FIRE algorithm. When $\Theta < 90^\circ$, $P$ is positive, and when $\Theta > 90^\circ$, $P$ is negative. Descent along the velocity direction with angle $\Theta$ is shown in Figure S1 (see Supplementary Information). Initially, at the start of the optimisation process, $\Theta$ is $0^\circ$. This happens because both the force and incoming velocity vectors are aligned with the negative gradient computed at the initial location. However, as we move forward, $\Theta$ gradually increases. 

At the minimum of the descent direction, $\Theta$ approaches $90^\circ$, showing that the force vector is orthogonal to velocity. As we move past the minimum, $\Theta$ increases until it hits $180^\circ$, indicating that the force vector is now anti-parallel to the velocity direction. A simple analysis as shown in Supplementary Information (see Table S3) suggests that, if an inexact line search is used, an algorithm selecting direction based on $\Theta$ may be an optimum one. As shown in Table S3, for $\Theta$ less than 90$^\circ$ and greater than 30$^\circ$, a direction taken with PR will get you closest to the minima. Similarly, for $\Theta$ greater 90$^\circ$ and less than 170$^\circ$, a direction taken with HS will take you closest to the minima. Based on this analysis, we tested several combinations of angle/direction in comparison to just taking single CG-like direction. And we find that using PR or FR direction when the angle is below 90$^\circ$, switching to HS between 90$^\circ$ and 120$^\circ$, and taking an SD direction beyond that yielded the best results. This methodology is used in our new algorithm to determine the modified velocity direction. 

\subsection{Adaptive Numerical Time-Step} 
We modify the time-step based on angle $\Theta$ in our new algorithm. When $\Theta < 90^\circ$, we increase the time-step $\Delta t$ by a factor of 1.1 and when $90^\circ < \Theta < 120^\circ$, that is, when we reach the region of minimum along a direction, we slow down by decreasing $\Delta t$ by a factor of 0.5.

As discussed earlier in Section 3.1, it is evident that because of overshooting, i.e. when the angle $\Theta$ is greater than 90$^\circ$, the efficiency of an algorithm decreases. If the overshooting is large, significant effort is required to come back. Therefore, to increase the efficiency of our algorithm, we incorporated a method where we return to the previous point and then move with half the time-step and velocity when $\Theta$ is greater than 120$^\circ$. Further, to overcome the problem of crawling, we do away with the latency time and the zeroing of velocities on overshooting. Instead, we reduce the time-step to half when $\Theta$ is greater than 90$^\circ$ and reduce the velocities to half when $\Theta$ is greater than 120$^\circ$. 

Because we are adaptively changing velocity and time-step, we term this algorithm as adaptively accelerated relaxation engine (AARE). When PR is used to determine direction for $\Theta$ less than $90^\circ$, we term the algorithm as ``AARE-PR'' and when FR is used, we term the algorithm as ``AARE-FR''. We elaborate on the performance of AARE with the original FIRE below (\textit{vide infra}). Here, we mention in passing that because of the above modifications, overshooting and crawling are avoided in AARE trajectories (see Figures \ref{fig:limitation}c and  \ref{fig:limitation}d).

\subsection{Algorithm}

The final AARE algorithm is as follows:
\begin{enumerate}[align=left]
  \item[\textbf{Step 1:}]
  Begin:
  \begin{itemize}
    \item Set AARE parameters: initial time-step $ \Delta t_{\text{start}}  = 0.1 \mathrm{fs}$; maximum time-step $ \Delta t_{\text{max}} = 10 \Delta t_{\text{start}}$; increment factor for time-step $ f_{\text{inc}} = 1.1$; decrement factor for time-step $f_{\text{dec}} = 0.5$.
    \item Initialize: time $t=0$; position $\mathbf{r} = \mathbf{r}_{\text{start}}$; velocity $\mathbf{v} = 0$; time-step $\Delta t = \Delta t_{\text{start}}$; iteration $k =1$. 
  \end{itemize}
  \item[\textbf{Step 2:}]
  Compute the forces $\mathbf{F}[\mathbf{r}(t)] = -\nabla U[\mathbf{r}(t)]$. Check for Convergence. If converged, exit the program.
 \item[\textbf{Step 3:}]
   If $k > 1$, compute $\Theta$: angle between $\mathbf{F}_{k}$  and $\mathbf{d}_{k-1}$
 \item[\textbf{Step 4:}]
   If $k > 1$, Compute $\beta_k$ : 
   \begin{itemize}
    \item If $\Theta < 90^\circ$: use PR or FR
    \item If $90^\circ < \Theta < 120^\circ$: use HS
    \item If  $\Theta > 120^\circ$: use SD (i.e., $\beta_k = 0$)
  \end{itemize}
 \item[\textbf{Step 5:}]
   Modify velocity : 
   \begin{itemize}
       \item If $k = 1$: $\mathbf{d}_k =  \mathbf{F}_{k}$
       \item If $k > 1$: $\mathbf{d}_k =  \mathbf{F}_{k} + \beta_k\mathbf{d}_{k-1}$
       \item Set $\mathbf{v} \rightarrow |\mathbf{v}| \cdot \frac{\mathbf{d}_k}{| \mathbf{d}_k |}$
   \end{itemize}
  \item[\textbf{Step 6:}] 
  For $k > 1$, update time-step $\Delta t$:
  \begin{itemize}
    \item If $\Theta < 90^\circ$: $\Delta t \rightarrow \min\left(\Delta t \cdot f_{\mathrm{inc}}, \Delta t_{\mathrm{max}}\right)$
    \item If $90^\circ < \theta < 120^\circ$: $\Delta t \rightarrow \Delta t \cdot f_{\mathrm{dec}}$
  \end{itemize}
 \item[\textbf{Step 7:}]
  Calculate $\mathbf{r}(t+\Delta{t})$, $\mathbf{v}(t+\Delta{t})$ using MD integrator 
 \item[\textbf{Step 8:}]
  Check for overshoot: 
  \begin{itemize}
    \item  Compute $\mathbf{F}_{k+1} = \mathbf{F}[\mathbf{r}(t + \Delta t)]$
    \item  Compute $\Theta$, the angle between $\mathbf{F}_{k+1}$  and $\mathbf{d}_{k}$
    \item if $\Theta > 120^\circ$,  return to $\mathbf{r}(t)$, $\mathbf{v} \rightarrow \frac{\mathbf{v}}{2}$ and $\Delta t \rightarrow \Delta t \cdot f_{\mathrm{dec}}$
    \item  Calculate $\mathbf{r}(t+\Delta{t})$, $\mathbf{v}(t+\Delta{t})$ using MD integrator 
  \end{itemize}
  \item[\textbf{Step 9:}]
  $ k = k + 1$, Go to Step 2. 
\end{enumerate}

In this work, we have used forward Euler method \cite{Atkinson1978} (for formula, see Supplementary Section S3) for integrating MD equations. Using a different integration scheme, like velocity verlet or semi-implicit Euler may improve the performance of our algorithm. As shown previously, this is true for the original FIRE algorithm.\cite{Shuang2019} We do not test these integrators in this work and can be taken up in future. Additionally, one can imagine applying the acceleration routine to the directions obtained from quasi-Newton methods instead of CG methods, which is an apt subject for future exploration. 

\section{Accelerated Conjugate-Gradient (Acc-CG)}
The AARE algorithms incorporate CG-like directions to guide the MD trajectory towards the minima, but they do not perform line search and instead change directions at each step. Another alternative approach is to use the traditional conjugate gradient (CG) algorithm itself instead of FIRE in optimization. In a traditional CG algorithm, a line search algorithm is used to determine the step length such that angle $\Theta$ is 90$^\circ$. Despite the effectiveness of line search in many scenarios, its requirement for repeated evaluations of the objective function hinders its application for computationally expensive functions. This underscores the need for an alternative approach to determining the appropriate step-size with fewer force evaluations, leading to the development of an algorithm that is more computationally efficient than FIRE.

Herein, we propose a CG algorithm that finds optimal step-sizes using an accelerated line search routine. This new algorithm is termed as accelerated CG (Acc-CG). The accelerated line-search combines the ideas of Newton's method and the secant method to find the step-sizes based on the angle $\Theta$. The force projection onto the search vector vanishes at minima, i.e.\ when $\Theta$ reaches $90^\circ$. Thus, our algorithm aims at finding step-size that brackets $\Theta$ between $80^\circ$ and $100^\circ$. The line-search is initiated in the new search direction using Newton's method.\cite{chongoptimization} In several cases, we find that this may lead to a very large initial step-size; forcing the algorithm to have instabilities. Therefore, in our algorithm, we add an additional constrain to limit the initial step-size using the step-size of the previous iteration. If the calculated step-size doesn't fall in between $80^\circ$ and $100^\circ$, and is less than $80^\circ$ we keep on doubling the step-size along the same search direction. In case there is overshooting, that is, $\Theta > 100^\circ$, the secant method \cite{chongoptimization} is used to interpolate and find the step size that brackets $\Theta$ between $80^\circ - 100^\circ$ interval. Once the step-size is determined, the next conjugate direction could be calculated using any CG formula. Here we have used the Polak-Ribi{\`e}re (PR) formula to find the new direction.

The accelerated CG (Acc-CG) algorithm is as follows:
\begin{enumerate}[align=left]
  \item[\textbf{Step 1:}]
  For iteration number $k =1$, start from the initial guess $\mathbf{r}_k$
  \item[\textbf{Step 2:}]
  Compute the forces $\mathbf{F}_k = -\nabla U[\mathbf{r}_k]$. Check for Convergence. If converged, exit the program.
   \item[\textbf{Step 3:}]
   Compute $\beta_k$ : 
   \begin{itemize}
    \item If $k = 1$: $\beta_k = 0$
    \item If $k > 1$: Polak-Ribi{\`e}re (PR) formula
  \end{itemize}
  \item[\textbf{Step 4:}]
   Compute new direction: 
   $\mathbf{d}_k = \mathbf{F}_k + \beta_k \mathbf{d}_{k-1}$
   \item[\textbf{Step 5:}]
   Evaluate the initial step-size $(\alpha)$ using Newton's method.
   \item[\textbf{Step 6:}]
   Calculate the new point: $\mathbf{r}_{k+1} = \mathbf{r}_k + \alpha \frac{\mathbf{d}_{k}}{\|\mathbf{d}_k\|}$
   \item[\textbf{Step 7:}]
  Compute the forces $\mathbf{F}_{k+1} = -\nabla U[\mathbf{r}_{k+1}]$
  \item[\textbf{Step 7:}]
  Compute $\Theta$: Angle between $\mathbf{F}_{k+1}$ and $\mathbf{d}_{k}$
  \begin{itemize}
    \item If $\Theta < 80^\circ$: $\alpha \rightarrow 2 \alpha$, $\mathbf{d}_{k+1} = \mathbf{d}_{k}$, $k = k +1 $, Go to Step 6.
    \item If $\Theta > 100^\circ$: Find new $\mathbf{r}_{k+1}$ using secant method, $k= k+1$, Go to Step 2.
    \item Else: $k = k+1$, Go to Step 3.
  \end{itemize}
\end{enumerate}
We assessed the efficiency of Acc-CG in comparison to traditional CG implementation in terms of number of force evaluations required for optimization of 2D test functions (see Table S4). In the traditional CG approach, we employed the Golden-section method to optimize the step size. Our results consistently showed that Acc-CG required fewer force evaluations to reach the desired minima compared to the traditional CG method. We also applied acceleration routines to the line-search with the traditional SD algorithm and observed a similar improvement in performance (see Table S4). One can also imagine applying the acceleration routines to the directions obtained from quasi-Newton methods, which may be done in future. 

\section{Application of Accelerated Algorithms}
We now discuss applications of AARE and accelerated CG algorithm on various 2D and 4D analytical functions.\cite{Andrei2008,Bongartz1995} We also illustrate the convergence of these algorithms for NEB pathways on simple 2D functions\cite{JONSSON1998} and HCN/CNH isomerization reaction; and compare their performance with the original FIRE algorithm. 

\subsection{Optimization on 2D and 4D Analytical Functions}
We begin by understanding the behaviour of AARE, Acc-CG and FIRE on several $two$-dimensional and $four$-dimensional functions. The mathematical form of these functions are given in Table S1 (see Supplementary Information). We access the performance of these algorithms by the number of force evaluations required to achieve the convergence criteria ($\|\mathbf{F}\| < 0.01$) (see Tables \ref{table:2D} and \ref{table:4D}). For all the cases studied herein, we find that AARE and Acc-CG outperforms FIRE. In some cases, the improvement achieved is more than 5 times.  The performance of AARE and Acc-CG is, however, comparable and we are not able to decide which is faster. For some cases, Acc-CG is faster than AARE and in others, AARE shows better performance than Acc-CG. For example, for the Himmelblau function, starting with (0,0), AARE-FR can achieve convergence to the minima in 25 force evaluations. On the other hand, Acc-CG takes 34 force evaluations to reach the minima with desired convergence criteria. However, for Quadratic QF1 function, Acc-CG reaches the minima in 9 force evaluations, whereas AARE-PR has to do an effort of 14 force evaluations. It is not a surprise that Acc-CG performs better for Quadratic QF1 and Perturbed Quadratic functions, because CG is designed to reach the minima in as many steps as the number of dimensions for a quadratic surface if an exact line-search is used for each step. This is typically not the case for a non-quadratic surface.

\begin{figure}[!h]
    \centering
    \includegraphics[width=1\linewidth]{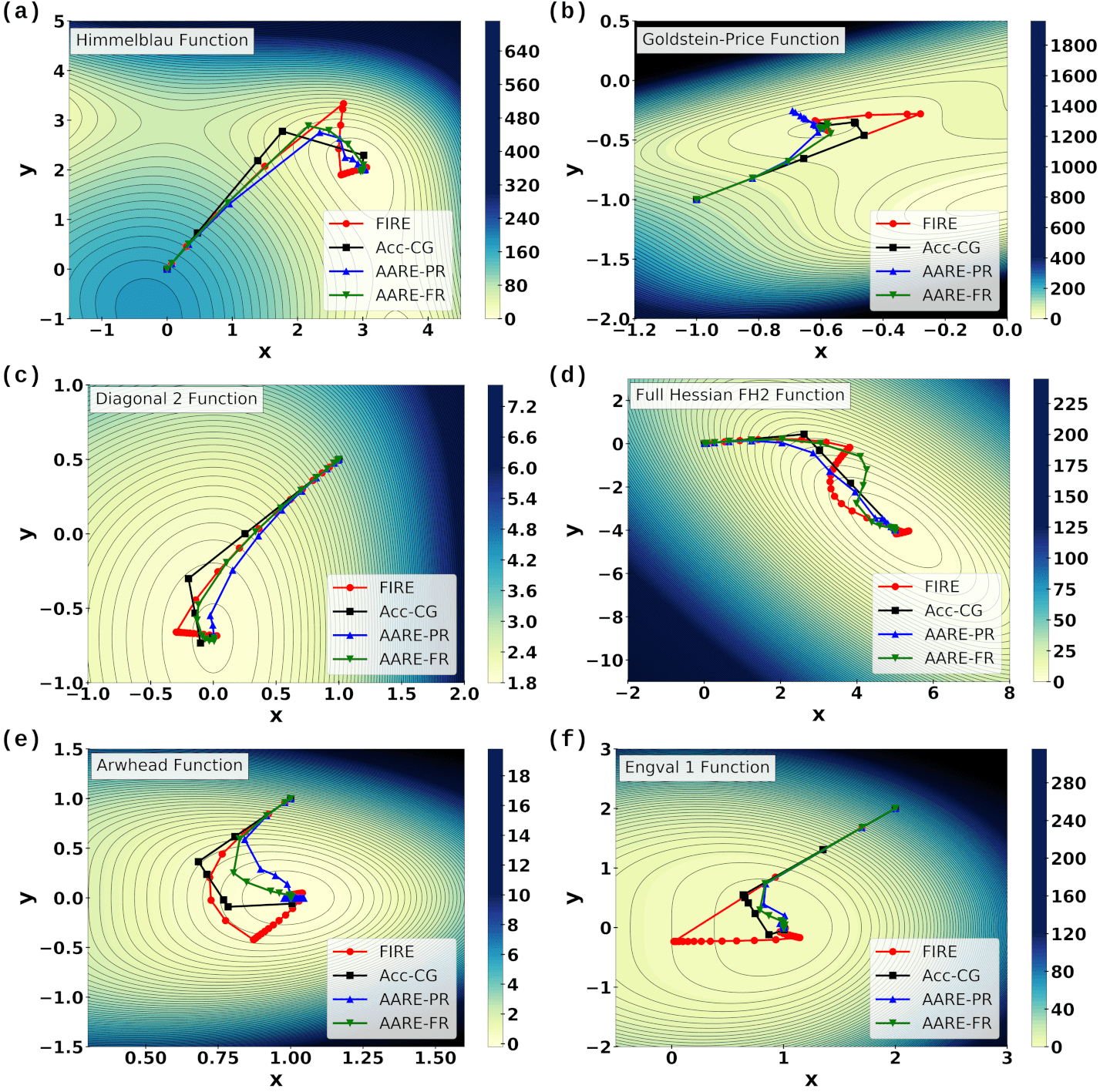}
     \caption{Trajectories of FIRE, ACC-CG, AARE-PR and AARE-FR on selected 2D test functions. The mathematical formulas for these functions are provided in Supplementary Information (Table S1). We used a convergence criteria of $\|\mathbf{F}\| < 0.01$ for these runs.}
    \label{fig:2d_contours}
\end{figure}

We now analyze the trajectories of FIRE, AARE and Acc-CG on selected functions plotted in Figure \ref{fig:2d_contours}. There are several interesting observations. First, large overshooting is avoided by both AARE and Acc-CG. Second, on overshooting, FIRE quenches the velocities to zero and crawls towards the minima before accelerating, which is not the case with  AARE and Acc-CG. 

\begin{figure}[!h]
    \centering
    \includegraphics[width=1\linewidth]{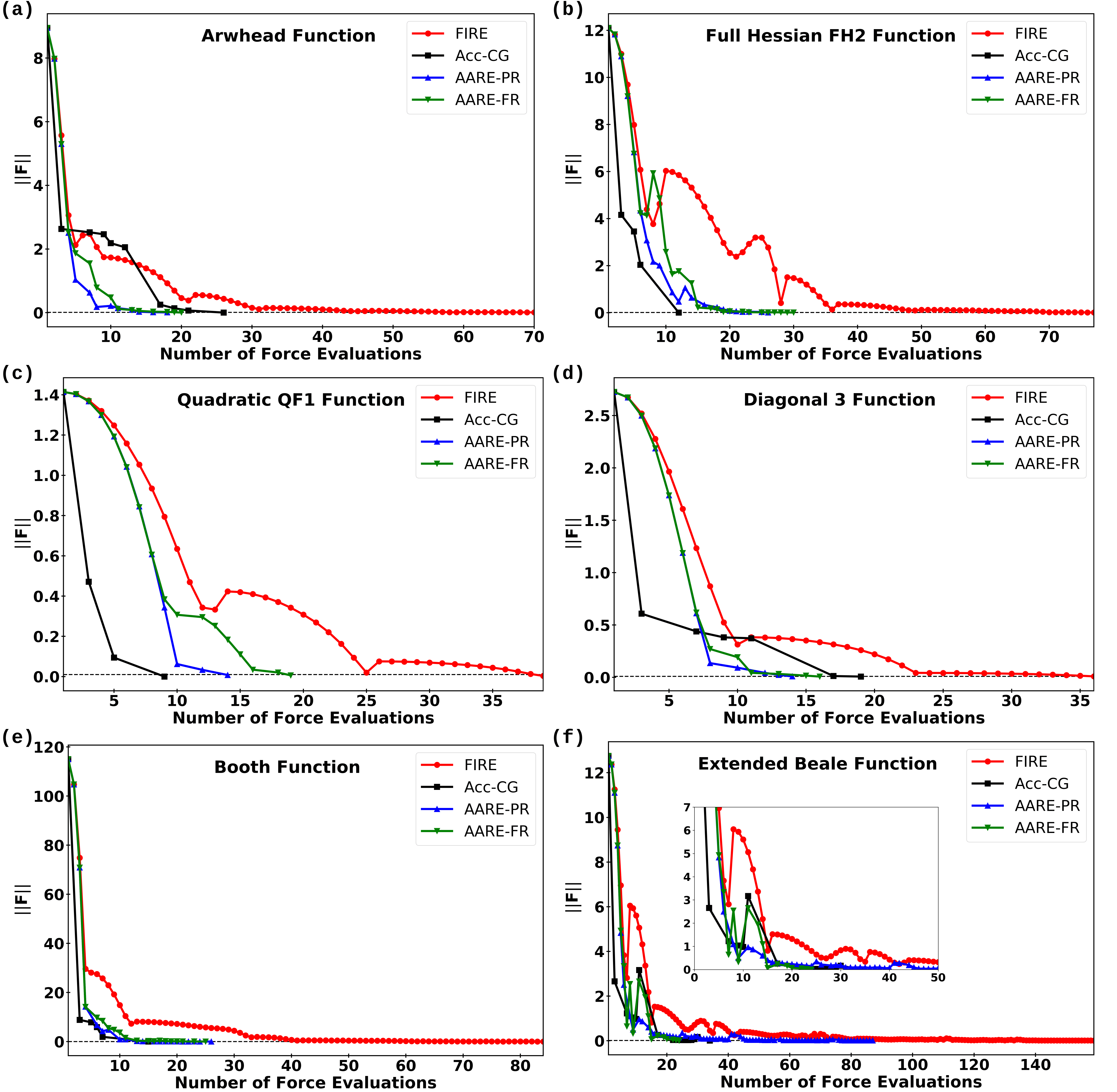}
     \caption{Euclidean norm of forces as a function of number of force evaluations for FIRE, Acc-CG, AARE-PR and AARE-FR. Mathematical formulas for the selected 2D test functions are provided in Supplementary Information (Table S1). The algorithm was terminated when $\|\mathbf{F}\| < 0.01$. }
    \label{fig:2d_normforce}
\end{figure}

We also plot the reduction in Euclidean norm of forces as a function of number of force evaluations in Figures \ref{fig:2d_normforce}. Examining the behaviour of norm of forces for Arwhead function and Full Hessian FH2 function (Figures \ref{fig:2d_normforce}a and \ref{fig:2d_normforce}b), we find that FIRE exhibits multiple overshoots. The overshoots results in reduction of integration time-step, which causes a slow convergence towards the minima.  However, both Acc-CG and AARE algorithms display a notable reduction in large overshoots compared to FIRE. Another key observation is that the initial reduction of forces is  faster in case of AARE and Acc-CG as compared to original FIRE. 

In the case of the Quadratic QF1 Function (Figure \ref{fig:2d_normforce}c), which has a near-quadratic nature, Acc-CG exhibits a quick reduction in forces compared to other algorithms. Overall, for most cases, we find that the initial drop in forces is large in Acc-CG, however it struggles near the minima, making its performance comparable to AARE. Similarly, for all cases, the initial reduction in forces is greater in AARE-PR as compared to AARE-FR. 

The Extended Beale and Rosenbrock serves as an example of slowly converging functions. In such cases, FIRE proves to be much slower compared to other algorithms, attributed to its tendency for multiple overshoots and velocity quenching (see Figures \ref{fig:2d_normforce}f). The adaptive acceleration mechanism employed in AARE and Acc-CG results in a more controlled decrease in forces, effectively mitigating the occurrence of large overshoots. Similar results are observed in 4D test functions as well. The reduction in the norm of forces with the number of force evaluations is given in Supplementary Information (Figure S2).

\begin{table}[!htbp]
\centering
\caption{Number of force evaluations needed by FIRE, Acc-CG, AARE-PR and AARE-FR for finding minima on 2D test functions. Also, included are the performance ratios relative to FIRE. The mathematical formulas for these functions are provided in Supplementary Information (Table S1). The iterative loop is terminated when $\|\mathbf{F}\| < 0.01$.}
\label{table:2D}
 \centering
 \large
 \resizebox{\textwidth}{!}{%
 \begin{tabular}{@{\extracolsep{\fill}}lc c c c c c c c c}
 \hline
 \hline
 \textbf {2D Test Functions} & \textbf {Starting Point} & \textbf{FIRE} & \textbf{Acc-CG} &\textbf{$ \frac{\textbf{FIRE}}{\textbf{Acc-CG}}$}& \textbf{AARE-PR} & \textbf{$ \frac{\textbf{FIRE}}{\textbf{AARE-PR}}$} & \textbf{AARE-FR} & \textbf{$ \frac{\textbf{FIRE}}{\textbf{AARE-FR}}$}\\ 
 \hline 
Himmelblau Function & (0,0) & 84 & 34 & 2.47 & 28 & 3 & 24 & 3.5 \\
$\text{Goldstein}-\text{Price Function}$ & (-1,-1) & 47 & 33 & 1.42 & 26 & 1.8 & 25 & 1.88\\
Extended Beale Function & (0,0) &  159 & 34 & 4.67 & 87 & 1.82 & 24 & 6.62\\
Rosenbrock Function & (-1.2,1) & 1565 & 324 & 4.83 & 951 & 1.64 & 217 & 7.21 \\
Hager Function & (1,1) & 26 & 12 & 2.17 & 13 & 2 & 14 & 1.86\\
Booth Function & (0,-5) & 84 & 15 & 5.6 & 26 & 3.23 & 25 & 3.36\\
Raydan 1 Function & (3,2) & 38 & 16 & 2.37 & 18 & 2.11 & 17 & 2.23\\
Extended Penalty Function & (1,2) & 32 & 30 & 1.07 & 19 & 1.68 & 17 & 1.88\\
Diagonal 1 Function & (0.5,0.5) & 13 & 10 & 1.3 & 10 & 1.3 & 10 & 1.3\\
Diagonal 2 Function & (1,0.5) & 41 & 15 & 2.73 & 16 & 2.56 & 18 & 2.28\\
Diagonal 3 Function & (1,1) & 36 & 19 & 1.89 & 14 & 2.57 & 16 & 2.25\\
Tridiagonal 1 Function & (2,2) & 49 & 22 & 2.23 & 14 & 3.5 & 16 & 3.06\\
Extended TET Function & (0.1,0.1) & 37 & 23 & 1.61 & 17 & 2.17 & 16 & 2.31\\
Generalized PSC1 Function & (3,0.1) & 40 & 23 & 1.74 & 11 & 3.63 & 6 & 6.67\\
Full Hessian FH2 Function & (0.01,0.01) & 77 & 12 & 6.41 & 26 & 2.96 & 30 & 2.57\\
Extended BD1 Function & (0.1,0.1) & 30 & 21 & 1.43 & 24 & 1.25 & 33 & 0.91\\
Extended Maratos Function & (1.1,0.1) & 1616 & 618 & 2.61 & 968 & 1.67 & 252 & 6.41\\
Quadratic QF1 Function & (1,1) & 30 & 9 & 3.33 & 14 & 2.14 & 19 & 1.58\\
Perturbed Quadratic Function & (0.5,0.5) & 36 & 10 & 3.6 & 18 & 2 & 18 & 2\\
Fletcher Function & (0,0) & 57 & 18 & 3.17 & 26 & 2.19 & 25 & 2.28 \\
Tridia Function & (1,1) & 54 & 10 & 5.4 & 20 & 2.7 & 19 & 2.84\\
Arwhead Function & (1,1) & 70 & 26 & 2.69 & 18 & 3.89 & 20 & 3.5\\
EG2 Function & (1,1) & 42 & 16 & 2.62 & 15 & 2.8 & 18 & 2.33\\
Liarwhd Function & (4,4) & 126 & 51 & 2.47 & 31 & 4.06 & 31 & 4.06 \\
Power Function & (1,1) & 53 & 14 & 3.78 & 18 & 2.94 & 22 & 2.41\\
Engval 1 Function & (2,2) & 67 & 26 & 2.57 & 21 & 3.19 & 16 & 4.18\\
\hline
\end{tabular}}
\end{table}

\begin{table}[!htbp]
\centering
\caption{Number of force evaluations needed by FIRE, Acc-CG, AARE-PR and AARE-FR for finding minima on 4D test functions. Also, included are the performance ratios relative to FIRE. The mathematical formulas for these functions are provided in the supplementary information (Table S1). Convergence criterion for optimization is set as $\|\mathbf{F}\| < 0.01$.}
\label{table:4D}
 \centering
 \large
 \resizebox{\textwidth}{!}{%
 \begin{tabular}{@{\extracolsep{\fill}}lc c c c c c c c c}
 \hline
 \hline
 \textbf {4D Test Functions} & \textbf {Starting Point} & \textbf{FIRE} & \textbf{Acc-CG} &\textbf{$ \frac{\textbf{FIRE}}{\textbf{Acc-CG}}$}& \textbf{AARE-PR} & \textbf{$ \frac{\textbf{FIRE}}{\textbf{AARE-PR}}$} & \textbf{AARE-FR} & \textbf{$ \frac{\textbf{FIRE}}{\textbf{AARE-FR}}$}\\ 
 \hline 
Himmelblau function & (1,1,1,1) & 88 & 29 & 3.03 & 23 & 3.82 & 32 & 2.75 \\
Extended Beale Function & (1,0.8,1,0.8) & 124 & 13 & 9.53 & 84 & 1.47 & 60 & 2.07 \\
Raydan 1 Function & (1,1,1,1) & 55 & 19 & 2.89 & 17 & 3.23 & 19 &2.89 \\
Extended Penalty function & (1,2,3,4) & 76 & 34 & 2.23 & 19 & 4 & 20 & 3.8 \\
Extended Trigonometric function & (0.2,0.2,0.2,0.2) & 56 & 28 & 2 & 23 & 2.43 & 29 & 1.93 \\
\hline
\end{tabular}}
\end{table}

\subsection{Optimization of NEB Pathway on 2D Test Functions}
We now turn our attention to accessing the performance of the new algorithms, in comparison to FIRE, for finding minimum energy path (MEP) connecting the two minima on the potential energy surface. One of the popular methods to obtain the minimum energy path (MEP) is the  Nudged Elastic Band method (NEB), which involves iteratively optimizing a series of connected images between energy minima towards the MEP. The images are linked together by springs, which creates an elastic band. The NEB force on the this elastic band is computed by summing the perpendicular component of the true force (i.e.\ the negative gradient of potential energy) and the parallel component of the spring force. Relaxing the band images by zeroing the NEB force thus satisfy  $-\nabla U[\mathbf{R_i}]_{\perp} = 0$ indicating their alignment with the MEP. The purpose of the springs is to keep the images well separated.  The images are relaxed using an optimization algorithm that iteratively adjusts the coordinates of the images along the path, ultimately nullifying the NEB forces on reaching MEP. 

We use the LEPS-I and LEPS-II models\cite{JONSSON1998} as test potential functions for the above purpose. Each of these functions have a first-order saddle-point connecting two minima. LEPS-I is a model potential for the reaction involving three atoms: AB reacts with C to form BC and A. The motion of the atoms is constrained to lie along a line. The reaction can be represented by a 2-D contour plot, where the $x$-coordinate represents the A-B distance and the $y$-coordinate represents the B-C distance. In LEPS-II model potential, the atoms A and C are fixed and B forms a bond with one of them in the minimum well. B atom is also coupled harmonically to the atom D. The potential function is represented as 2-dimensional with  $x$-coordinate as A-B distance and $y$-coordinate as B-D distance. The functional form of these potentials are given in Supplementary Information (see section S2). 

The NEB calculations were set up by creating a guess for MEP using linear interpolation between the two minima. A total of \emph{twelve} images including the minima were used, which were connected by springs having $k_{\text{spring}} = 1$. The tangent at the image was estimated using the improved-tangent NEB (IT-NEB) method.\cite{Henkelman2000} The number of force evaluations required to reach the desired accuracy are given in Table \ref{table:MEP}. We calculate the computational cost as the number of force evaluations because if we were to perform calculations with DFT, a single point calculation will provide both energy and forces at a similar computational cost. For a NEB band of 10 images, we count the number force evaluations as 10. Note that the forces at the minima are evaluated once during the entire NEB because they are fixed. 

For both LEPS-I and LEPS-II, we find that both AARE and Acc-CG perform better than FIRE. AARE-FR, however, reaches the minima in fewest of force evaluations. The progress of AARE-FR on LEPS-I is shown in  Figure \ref{fig:LEPSI}a and the rest of the trajectories are shown in Figures S3 and S4 of Supplementary Information. The reduction in the norm of forces with the number of force evaluations is shown in Figure \ref{fig:LEPSI}b. A similar trend emerges as seen in optimization on 2D test functions. Notably, both the Acc-CG and AARE algorithms exhibit a more rapid decrease in NEB forces compared to FIRE. These new algorithms effectively reduce overshooting in forces, though a notable exception is observed with Acc-CG, possibly due to the occurrence of kinks during NEB iterations. Nevertheless, as NEB progresses, both Acc-CG and AARE demonstrate stable and accelerated force reduction. AARE-PR initially outperforms all the algorithms in force reduction, but near the minima AARE-FR emerges as the superior choice among all the tested algorithms. One could also think of combining AARE-PR followed by AARE-FR for an optimal performance. This combination could potentially yield a faster initial decrease in forces while ensuring stable reduction as the NEB approaches the MEP---an apt subject for future exploration.

\begin{table}[!htbp]
\centering
\caption{Number of force evaluations taken by the new algorithms: AARE-PR, AARE-FR and Acc-CG; relative to FIRE for finding MEP. The convergence criterion was set as $\|\mathbf{F}\| < 0.01$, for analytical functions and $\|\mathbf{F}\| < 0.01 \text{ eV/\AA}$ for HCN isomerization. For LEPS-I and LEPS-II, the discretized pathway had 12 images and a $k_{spring} = 1$ was used. For HCN/CNH isomerization, the discretized pathway had 11 images and a $k_{spring} = 1 \text{ eV/\AA} ^2$ was used.}
\label{table:MEP}
 \centering
 \large
 \resizebox{\textwidth}{!}{%
 \begin{tabular}{@{\extracolsep{\fill}}lc c c c c c c c c}
 \hline
 \hline
\textbf{PES} & \textbf{FIRE} & \textbf{Acc-CG} &\textbf{$ \frac{\textbf{FIRE}}{\textbf{Acc-CG}}$}& \textbf{AARE-PR} & \textbf{$ \frac{\textbf{FIRE}}{\textbf{AARE-PR}}$} & \textbf{AARE-FR} & \textbf{$ \frac{\textbf{FIRE}}{\textbf{AARE-FR}}$}\\
\hline
LEPS I & 1782 & 1242 & 1.43 & 1252 & 1.42 & 982 & 1.81 \\
LEPS II & 1352 & 952 & 1.42 & 902 & 1.5 & 702 & 1.93 \\
HCN Isomerization & 1343 & 1064 & 1.26 & 2540 & 0.53 & 929 & 1.46 \\
\hline
\end{tabular}}
\end{table}

\begin{figure}[!htbp]
           \centering
        \includegraphics[width=\linewidth]{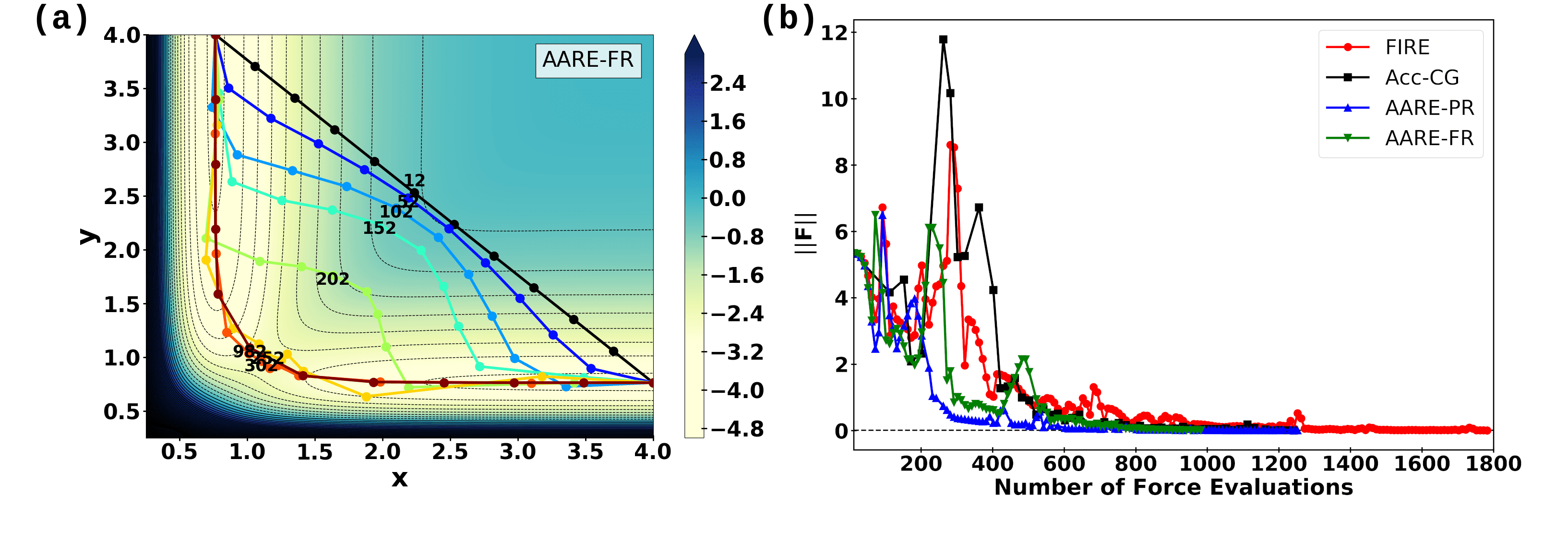}
\caption{\textbf{(a)} Illustrates NEB pathways for the LEPS-I potential  using AARE-FR as an optimizer. The black curve represents the initial discretized pathway. The numbers on each intermediate pathway represents the number of force evaluations required to reach at that point. The final MEP, achieved after 982 force evaluations, is shown in red. \textbf{(b)} Euclidean norm of forces as function of the number of force evaluations for FIRE, Acc-CG, AARE-PR, and AARE-FR during an optimization run on LEPS-I potential. The mathematical formula for the LEPS-I is provided in Supplementary Information (Section S2). We use the convergence criterion as $\|\mathbf{F}\| < 0.01$ and a $k_{\text{spring}} = 1$ for these runs.}
    \label{fig:LEPSI}
\end{figure}

\subsection{HCN/CNH Isomerization}
In this section, we use HCN/CNH isomerization reaction as an illustrative vehicle to access the performance of the three algorithms. HCN is an intermediate during combustion of hydrocarbons in air at high temperatures, and therefore, plays a critical role in studying its chemistries. The system has a total of nine degrees-of-freedom (DOF); out of which five are the translational and rotational degrees of freedom and rest four are the vibrational DOF. To eliminate the rotational and translational DOF, we have fixed the carbon atom and allowed nitrogen atom to move only in $x$-direction. Because of symmetry, hydrogen was constrained to move only in the $xy$-plane. This reduces the problem to three DOF. 

The forces and the ground-state energy, for a fixed geometry of (H)CN, was computed by performing spin-polarized density functional theory (DFT) calculations using Gaussian16 software package.\cite{g16} The effective potential in the DFT was obtained using the B3LYP hybrid functional and the wave functions were approximated using a double-zeta basis-set [6-31+G(d)]. \cite{Petersson1988,Petersson1991,Becke1993} No symmetry constraints were applied to the orbitals. The geometrical parameters of the critical points on the PES are listed in Table S5, which are a good match with previous calculations. The product, CNH, is 0.64 eV less stable than the reactant and the isomerization reaction proceeds with an activation barrier of 2.12 eV.

The minimum energy path (MEP) was obtained using the NEB method, starting with an initial guess consisting of 11 images including the reactant and the product. This is a typical case where linear interpolation between the  reactant and the product configuration would have generated images with atoms overlapping with each other. This would cause DFT calculations in Gaussian16 to fail. To circumvent this issue, we created an image perpendicular to C-N bond at a distance 1.5 \AA\ from the the center of C$\equiv$N. The other images were created by linearly interpolating between this image and the reactant/product. There are other methods of generating initial guess of images,\cite{Halgren1977,Smidstrup2014} for example linear-synchronous-transit (LST) method. The use of these methods will not change the qualitative nature of the findings in this work. The images were joined by springs of $k_{\text{spring}} = 1$ eV/\AA$^2$. 

After setting the initial guess for NEB, we performed the calculations using FIRE, Acc-CG and AARE algorithms. Table \ref{table:MEP} summarizes the number of force evaluations required by each algorithm. Upon comparing their performances, we found that AARE-FR performed the best, followed by Acc-CG, FIRE, and AARE-PR. 

\begin{figure}[!htbp]
           \centering
        \includegraphics[width=0.7\linewidth]{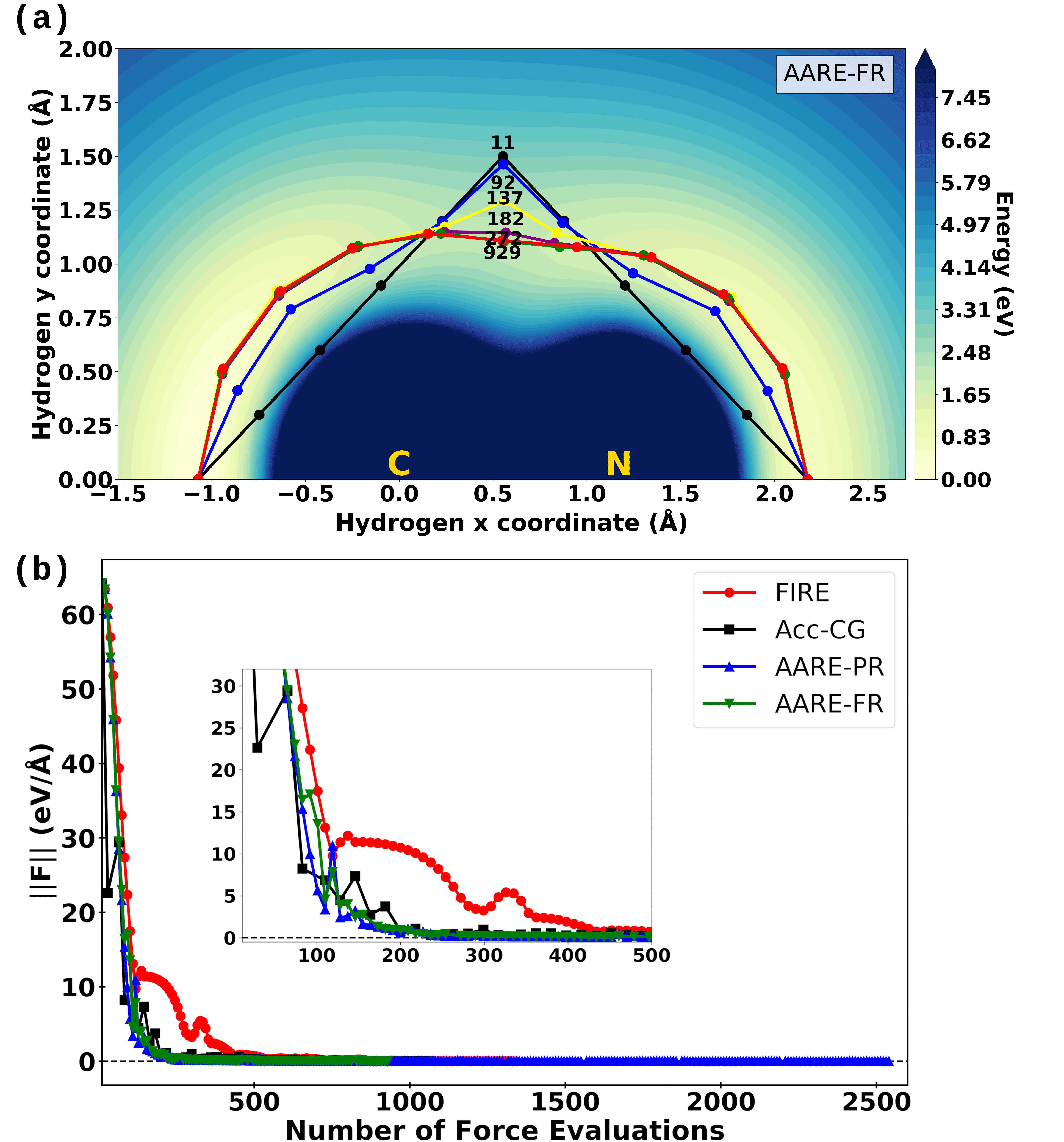}
\caption{\textbf{(a)} Discretized pathways during an optimization of the NEB calculation using AARE-FR. The starting band is shown in black.  The numbers on the pathways indicate the number of force evaluations needed by AARE-FR. The final MEP (red line) is obtained after 929 force evaluations. \textbf{(b)} Euclidean norm of forces as a function of number of force evaluations using FIRE, Acc-CG, AARE-PR, and AARE-FR. We use a convergence criterion of $|\mathbf{F}| < 0.01 \text{ eV/\AA}$ with $k_{\text{spring}} = 1 \text{ eV/\AA}^2$. }
    \label{fig:HCN}
\end{figure}

As previously discussed, the HCN/CNH isomerization is a three-dimensional problem. For simplicity and to visualize the NEB pathways, we plotted 2-dimensional contours of the reaction pathway in Figure \ref{fig:HCN}. The $x$-axis represents the $x$-coordinate of the hydrogen atom, and the $y$-axis represents the $y$-coordinate of the hydrogen atom. The energy in the plot corresponds to the most stable C-N distance for a particular H coordinates. In these calculations, we fixed the carbon atom and allowed the nitrogen atom to move only along the $x$-axis. On this plot, we displayed the NEB pathways optimized using AARE-FR (Figure \ref{fig:HCN}a). The black line indicates the initial guess of images. 

We also analyzed the reduction in the norm of forces with respect to the number of force evaluations, as shown in Figure \ref{fig:HCN}b. Interestingly, we found that AARE-PR had the fastest initial convergence among all the algorithms but slowed near the minima. This is consistent with the findings for 2D analytical test functions.

\section{Conclusion}
In this work, we developed two new optimization algorithms: the Adaptively Accelerated Relaxation Engine (AARE) and the Accelerated Conjugate-Gradient (Acc-CG). AARE is an MD-based optimization scheme that steers the trajectory in a conjugate-gradient-like direction, while Acc-CG is an enhanced version of the traditional conjugate-gradient method, integrating Newton and secant methods to optimize step size along each search direction. Both algorithms address the limitations of the widely used FIRE optimization scheme, particularly its reliance on empirical parameter and issues with convergence speed and overshooting, by adaptively selecting directions based on local gradient information.

A key in these algorithms is the use of the conjugate-gradient direction to guide the search for minima, eliminating the need for empirical mixing parameter. Additionally, both algorithms analyze the local nature of the function by calculating the angle $\Theta$, similar to the power factor in FIRE. This approach allows for smoother trajectory adjustments, thereby reducing overshooting and crawling. Two variations of AARE are proposed: AARE-PR and AARE-FR, which use the Polak-Ribi{\`e}re and Fletcher-Reeves formulas; respectively, for calculating the CG directions. AARE also includes a mechanism to prevent velocities from becoming zero during uphill motion, thereby addressing the crawling problem observed in FIRE. In Acc-CG, the new step-size calculation method replaces the computationally expensive line search, and overshoots are controlled using the secant method to return the system to the optimal step length.

The performance of these algorithms was tested on 2D and 4D test functions for identifying minima. Our results show that both AARE and Acc-CG achieve a faster initial reduction in the norm of forces compared to FIRE. Acc-CG performed exceptionally well on nearly-quadratic problems, benefiting from the CG algorithm's convergence efficiency on such surfaces. Both AARE algorithms exhibited similar behavior, with AARE-PR showing a faster initial decrease in force than AARE-FR. However, AARE-PR required more force evaluations when approaching the minima, resulting in similar computational costs for both variations. On slow-converging functions, such as the Rosenbrock and Extended Beale functions, AARE-FR demonstrated significantly better convergence than AARE-PR.

Subsequently, these optimization methods were integrated with the Nudged Elastic Band (NEB) method to find the Minimum Energy Path (MEP). Initial testing was conducted on analytical 2D potential energy functions, such as LEPS-I and LEPS-II, before extending these strategies to determine the MEP for HCN/CNH isomerization. In all cases, the new algorithms proved more computationally efficient than FIRE, with AARE-FR consistently showing better force reduction.

It is important to note that although AARE and Acc-CG methods are tailored for NEB problems, they are versatile optimization algorithms for diverse applications.

\section*{Supporting Information}
Test functions for unconstrained optimization; mathematical form of analytical potential energy functions LEPS-I and LEPS-II; forward Euler integrator; selected geometrical parameters of HCN, CNH and transition state connecting the two minima; optimised value of $\alpha_{\text{start}}$ for optimisation of 2D test functions; change in Steepest-Descent, Hestenes-Stiefel, Polak-Ribi{\`e}re and Flecther-Reeves direction vectors while traversing along the search direction on a quadratic potential; euclidean distance calculated between the minima and the line search point by following Steepest-Descent, Hestenes$-$Stiefel, Polak-Ribi{\`e}re and Flecther-Reeves direction vectors while traversing along the search direction on a quadratic potential; number of force evaluations needed by Steepest-Descent and Conjugate-gradient algorithms, with step size calculated using Golden-section-based line search and Newton-Secant acceleration routine for finding minima on 2D test functions; euclidean norm of forces as a function of number of force evaluations for optimisation of 4D test functions, NEB pathways for LEPS-I and LEPS-II potentials, euclidean norm of forces as function of number of force evaluations for optimization on LEPS-II potential, NEB pathways and movie of HCN/CNH isomerization.

\section*{Conflicts of interest}
There are no conflicts to declare.

\section*{Acknowledgements}
S. L. S.\ is thankful for the financial support from the Prime Minister's Research Fellowship scheme, Govt.\ of India.
V.A.\ acknowledges the financial support from Anusandhan National Research Foundation (ANRF), Department of Science and Technology, Govt.\ of India (DST grant no.\ CRG/2023/000623). The authors also acknowledge the support of DST, Govt.\ of India, for the High-Performance Computing (HPC2013 and Paramsanganak) facility  at IIT Kanpur. 

\bibliography{AARE_AccCG} 

\end{document}